\documentclass[12pt]{iopart}

\usepackage{soul}
\RequirePackage{graphicx}
\usepackage{latexsym}
\expandafter\let\csname equation*\endcsname\relax
\usepackage{xcolor}
\expandafter\let\csname endequation*\endcsname\relax
\usepackage{amssymb}
\usepackage{amsmath} 
\usepackage{todonotes}
\usepackage{caption}
\usepackage{subcaption}
\usepackage{orcidlink}

\begin{document}

\title{$q$-Deformed Quantum Mechanics and the Thermodynamics of Black Hole/White Hole Spectral pair}

\author{S. Jalalzadeh$^{1,2,3}$\footnote{Author to whom any correspondence should be addressed} \orcidlink{0000-0003-4854-29600000-0003-4854-2960},  R. Jalalzadeh$^4$ \orcidlink{0000-0002-6110-3981}, and H. Moradpour$^4$ \orcidlink{0000-0003-0941-8422}}
\address{$^1$ Izmir Institute of Technology, Department of Physics, Urla, 35430, Izmir, Türkiye}
\address{$^2$ Center for Theoretical Physics, Khazar University, 41 Mehseti Street, Baku, AZ1096, Azerbaijan}
\address{$^3$ Department of Physics, Dogus University, Dudullu-Ümraniye, 34775 Istanbul, Türkiye}
 \address{$^4$ Research Institute for Astronomy and Astrophysics of Maragha (RIAAM),
 University of Maragheh, P. O. Box 55136-553, Maragheh, Iran}
 \ead{shahramjalalzadeh@iyte.edu.tr}

 \vspace{10pt}
\begin{indented}
\item[]
\end{indented}

\begin{abstract}
In this work, we investigate the thermodynamics of Schwarzschild black and white holes within a $q$-deformed Wheeler--DeWitt framework. By introducing a $q$-deformed Heisenberg--Weyl algebra at a root of unity, we derive a finite-dimensional Hilbert space, a bounded mass spectrum, and an adiabatic invariant leading to a bounded entropy-mass relation. The deformation results in a universal logarithmic correction, as well as a minimum temperature and a maximum entropy that matches the de Sitter bound. Also, we examine the interpretation of a cold remnant, which is dynamically stable because its radiation rate approaches zero, even though its heat capacity remains negative. We also explore the holographic implications of this limited entropy. 
Our results thus provide a consistent semiclassical picture, where quantum deformation naturally introduces an entropy bound, avoids divergences at the final evaporation stage, and suggests a smooth transition from quantum gravity to cosmology.
\end{abstract}

%
%
%
%

\section{Introduction} 
Black holes (BH) have made significant contributions to advancements in theoretical physics.  Moreover, in recent decades, numerous studies have revealed the presence of massive and supermassive BHs in the centers of typical galaxies.  The recent observations of gravitational waves by the LIGO and VIRGO collaborations \cite{LIGOScientific:2016aoc, LIGOScientific:2016sjg, LIGOScientific:2018jsj} from the merger of binary astrophysical BHs have unequivocally confirmed the existence of BHs in the universe.  In theoretical physics, BH phenomena are crucial for comprehending quantum gravity, a subject that has been widely investigated over the last several decades.

 {Like atoms, BHs can be assigned discrete quantum levels in semiclassical and minisuperspace approaches; unlike atoms, however, BHs radiate thermally through Hawking emission.}  One of the primary goals of the quantum theory of BHs is to predict their emission spectra.  The radiation spectrum, in turn, is determined by the spectra of BHs' classical observables. When a hole emits a quantum of BH radiation, it transitions from one classical observable eigenstate to the next. In a thought experiment, Bekenstein conducted a detailed calculation of the smallest increase in the horizon area resulting from the absorption of a particle by a Kerr--Newman BH \cite{Bekenstein:1973ur}.  He has convincingly shown that this specific rise in the horizon area represents a constant value, staying invariant and unaffected by the characteristics that define the BH, namely its mass, charge, and angular momentum. In 1974, Bekenstein \cite{1974NCimL} proposed, using the Bohr--Sommerfeld quantization condition, that the horizon area of a Kerr--Newman BH is quantized, with the horizon area in Planck area units being proportional to an integer. The quantization of the BH has been the focus of a lot of work \cite{Barvinsky:2000gf, Kuchar:1994zk, Ashtekar:2005qt, Jalalzadeh:2011yp, Gambini:2013ooa, Gambini:2013hna, RodriguesAlmeida:2021kck, Gambini:2023wnc, Jalalzadeh:2026ttf, Jalalzadeh:2026ssq}.  Vaz and Witten \cite{Vaz:1998gm} determined the mass eigenstate and eigenvalue of a Schwarzschild BH by solving the Wheeler--DeWitt (WDW) equation within the framework of canonical quantum gravity. The authors of \cite{Bina:2010ir} analyzed the effects of the generalized uncertainty principle (GUP) on canonical quantum gravity of BHs. Also, the thermodynamics of fractional-fractal BHs is studied in the context of the fractional WDW equation in Refs. \cite{Jalalzadeh:2021gtq, Jalalzadeh:2025uuv}.
Most of these articles support the idea that the entropy of BHs is quantized, i.e.,
\begin{equation}\label{Ent}
    S_\text{BH}=\gamma n,~~~~n~\text{is a large integer number},
\end{equation}
where $\gamma$ is a pure number of order one.
This shows that the entropy spectrum is quantized and equidistant for a spherically symmetric static
BH, i.e., $\Delta S_\text{BH}=\gamma$.

Historically, quantum groups have arisen from investigations into quantum integrable models, employing quantum inverse scattering techniques, which resulted in the deformation of classical matrix groups and their associated structures.  Lie algebras as discussed in the works of Kulish et al. \cite{kulish1983quantum}, Sklyanin \cite{sklyanin1982some}, and Faddeev \cite{faddeev2016liouville}.  It has been established that quantum groups assume a significant role in various domains, including quantum integrable systems \cite{chari1995guide}, conformal field theory \cite{oh1992conformal}, knot theory \cite{kauffman2007q}, solvable lattice models \cite{foda1994vertex}, topological quantum computations \cite{nayak2008non}, molecular spectroscopy \cite{chang1991q}, quantum gravity \cite{smolin1995linking,major1996quantum,dupuis2013quantum,dittrich2014quantum,rovelli2015compact, Jalalzadeh:2022rxx}, and quantum cosmology \cite{Jalalzadeh:2023mzw, Jalalzadeh:2017jdo}.

In the article \cite{Jalalzadeh:2023mzw}, the author used the $q$-deformed Heisenberg--Weyl algebra to find the entropy of a Schwarzschild BH. Briefly, they used the relation between the event horizon of a hole, $A=16\pi M^2G^2$, the relation of the entropy of the BH with the horizon area, $S_\text{BH}=A/(4G)$, with the $q$-deformed mass spectrum of the BH to obtain the spectrum of the entropy. They obtained a new $q$-deformed entropy given by the relation $S_q=\pi\sin(\frac{\pi}{\mathfrak N}(n+\frac{1}{2}))/\sin(\frac{\pi}{2\mathfrak N})$, where $n$ is an integer and $\mathfrak N$ is the $q$-deformation parameter. Although this relation reduces to Eq. (\ref{Ent}) for the c-number limit, i.e., {$\mathfrak N\rightarrow\infty$}, it is clear that it is not consistent with relation (\ref{Ent}). In this article, we show that the meticulous examination of calculations leads us to a similar expression as (\ref{Ent}) for $q$-deformed BHs. In addition, we obtained the temperature and heat capacity of the resulting $q$-deformed pair.

The structure of our paper is outlined as follows.  The next section is a concise summary of the $q$-deformed WDW equation related to a Schwarzschild black hole-white hole (WH) pair.  In section 3, we rederive the entropy, temperature, and heat capacity pertinent to the model.  The final section presents the key findings of our study. {Note that in our work, the ‘BH/WH pair’ denotes the two monotonic branches of a single finite $q$-deformed mass spectrum, rather than a literal astrophysical binary system.}


\section{$q$-deformed Schwarzschild quantum black hole}

{Birkhoff's theorem implies that the Schwarzschild mass is the only diffeomorphism-invariant parameter of the vacuum solution. After reducing the spherically symmetric gravitational system to its physical minisuperspace degree of freedom, the WDW equation can be mapped to a one-dimensional harmonic-oscillator eigenvalue problem. In this representation, \(x\) denotes the dimensionless canonical minisuperspace variable of the reduced Schwarzschild geometry, while the corresponding eigenvalue is proportional to \(M^2\). For completeness, a brief derivation of this equation is provided in \ref{canon}.}

The WDW equation of Schwarzschild geometry can be expressed as \cite{Jalalzadeh:2022rxx}:
 \begin{equation}
    \label{WDW1}
    -\frac{1}{2}\frac{\mathrm d^2\psi(x)}{\mathrm dx^2}+\frac{1}{2}x^2\psi(x)=\frac{2M^2}{m_\text{P}^2}\psi(x),
\end{equation}
where $\psi(x)$ is the wavefunction of the BH, and $m_\text{P}=1/\sqrt{G}$ is the Planck mass in
natural units, i.e., $\hbar= c = k_B = 1$. Upon the use of (\ref{WDW1}),  the  mass spectrum $M_n$ is
\begin{equation}\label{1-10}
M_n=\frac{m_\text{P}}{\sqrt{2}}\sqrt{n+\frac{1}{2}},~~~~~n=0,1,2,...~~.
\end{equation}

To effectively develop the quantum deformation of the above WDW equation, following Ref. \cite{Jalalzadeh:2022rxx}, we utilize the Heisenberg--Weyl algebra linked to the structure of the non-semisimple Lie algebra $h_4$, which includes four generators $\{a_+, a_-, N, 1\}$, where $1$ is the multiplicative identity, {$N=a^\dagger a$ is the number operator}, and $a_-,a_+$  are the usual annihilation and creation operators with usual commutation relations in quantum mechanics. 
 Consequently, the WDW equation (\ref{WDW1}) turns to: 
  \begin{equation}\label{3-12nona} 
  (a_+a_-+a_- a_+)|n\rangle=\frac{2M^2}{m_\text{P}^2}|n\rangle, 
  \end{equation}
where $|n\rangle$ is the mass eigenstate, and $\psi_n(x)=\langle x|n\rangle$.

The quantum Heisenberg--Weyl algebra, $U_q
(h_4)$, at the root of unity, is a $q$-deformation of the Heisenberg--Weyl algebra, which is a
associative unital \cite{Chaichian, Chaichian2} $\mathbb C(q)$-algebra with generators
$\{a_+,a_-,q^{\pm N/2}\}$. {Note that in the undeformed oscillator representation, the operator \(a^\dagger a\) acts as the usual number operator on the Fock basis. In the \(q\)-deformed algebra, however, \(N\) is introduced as an independent generator satisfying the defining commutation relations of the deformed Heisenberg--Weyl algebra. Although it reduces to the usual number operator role in the undeformed limit, it should not be identified \textit{a priori} with the operator \(a^\dagger a\) \cite{Chaichian}.} The generators satisfy the following quantum deformed ($q$-deformed) commutation relations \cite{Chaichian}
\begin{equation}\label{new4}
a_-a_+-q^{\frac{1}{2}}a_+a_-=q^{\frac{N}{2}},~~~[N,a_\pm]=\pm a_\pm,~~~
a_\pm^\dagger=a_\mp,~~~N^\dagger=N,~~~|q|=1,
\end{equation}
where $q$ is a primitive root of unity, i.e.,
$q:=\exp\left(\frac{2\pi i}{\mathfrak{N}}\right)$,
 $\mathfrak{N}$ is a natural number, and $\mathfrak{N}\geq2$.

The natural length scale in quantum gravity is the Planck length. Therefore, \( \mathfrak N \) can be expressed as a function of the gravitational constant \( G \) or the square of the Planck length: \( \mathfrak N = \mathfrak N(l_\text{P}^2) \). In the classical gravity limit, where \( l_\text{P} \) approaches zero, \( \mathfrak N \) tends to infinity, indicating a transition to classical gravity. This shows the $q$-deformation is a quantum gravity effect, with \( \mathfrak N \propto 1/l_\text{P}^2 \). 
To maintain \( \mathfrak N \) as a dimensionless parameter, another length scale is needed, where the ratio of the Planck length to this new length defines the $q$-deformation parameter: \( \mathfrak N = L_q^2 / l_\text{P}^2 \).
{The parameter \(L_q\) may be interpreted as an infrared length scale associated with the finite-dimensional \(q\)-deformed representation. At the algebraic level, the key input is the finiteness of the representation dimension \(N\). The identification \(N=L_q^2/l_P^2\) is an additional physical parametrization that allows one to associate the finite spectrum with a large-scale cutoff.}
 The classical gravity limit is reached by setting \( L_q \rightarrow \infty \). Also, this deformation parameter provides a holographic view of quantum mechanics \cite{Jalalzadeh:2021oxi}. Specifically, the Hilbert space of \( q \)-deformed neutral hydrogen gas in de Sitter space satisfies the strong holographic bound with this parameter.

One can easily  show
that the first two relations (\ref{new4}) are
actually equivalent to the following relations
\begin{equation}\label{new111}
a_+a_-=[N],~~~a_-a_+=[N+1],
\end{equation}
where 
\begin{equation}\label{number}
[x]:=\frac{q^\frac{x}{2}-q^\frac{-x}{2}}{q^\frac{1}{2}-q^\frac{-1}{2}}=\frac{\sin\left(\frac{\pi x}{\mathfrak{N}}\right)}{\sin\left(\frac{\pi}{\mathfrak{N}}\right)}.
\end{equation}
The $q$-deformed creation and annihilation operators act on the mass states by
\begin{equation}\label{3-10non}
\begin{split}
&a_+|n\rangle=\sqrt{[n+1]}|n+1\rangle,~~~
a_-|n\rangle=\sqrt{[n]}|n-1\rangle,~~~
 N|n\rangle=n|n\rangle,\\
& a_+|\mathfrak N\rangle=0,~~~~n=0,1,...,\mathfrak N-1.
 \end{split}
\end{equation}
Now, like the ordinary Fock space of the harmonic oscillator, we can construct the representation of the $ U_q(h_4)$ in the $q$-deformed Fock space spanned by normalized eigenstates $|n\rangle$
\begin{equation}\label{3-8non}
|n\rangle=\frac{1}{\sqrt{[n]!}}{a}_+^n|0\rangle,~~~~n=0,1,...,\mathfrak N-1,
\end{equation}
where the q-factorial is defined by $[n]!:=\prod_{m=1}^n[m]$. 

The operator
\begin{equation}\label{3-12non}
\frac{2M^2}{m^2_\text{P}}=a_+a_-+a_- a_+=[N+1]+[N],
\end{equation}
is the $q$-analogue of the mass operator defined in (\ref{3-12nona}). 

Consequently, the above quantum deformation of the Schwarzschild geometry gives us the following eigenvalues for the $q$-deformed mass \cite{Jalalzadeh:2022rxx}
\begin{equation}\label{3-19non1}
M_n=\frac{m_\text{P}}{2}\sqrt{\frac{\sin(\frac{\pi}{\mathfrak{N}}(n+\frac{1}{2}))}{\sin(\frac{\pi}{2\mathfrak{N}})}},~~~~~n=0,...,\mathfrak{N}-1.
\end{equation}
Note that for $\mathfrak{N}\rightarrow\infty$ the earlier eigenvalues will reduce to
(\ref{1-10}). Also, there is a two-fold degeneracy at the mass eigenvalues.  The mass of the ground state $n=0$, as well as the state $n=\mathfrak N-1$, are $M_{0}=M_{\mathfrak N-1}=\frac{m_\text{P}}{2}$. These show that the ground state mass is not deformed, and its value is the same as the non-deformed spectrum obtained in (\ref{1-10}). Besides, the spectrum is bounded in which the most excited state, $n=\mathfrak N-1$, has the same mass and area as the ground state.

\begin{figure}
     \centering
     \begin{subfigure}[b]{0.4\textwidth}
         \centering
         \includegraphics[width=\textwidth]{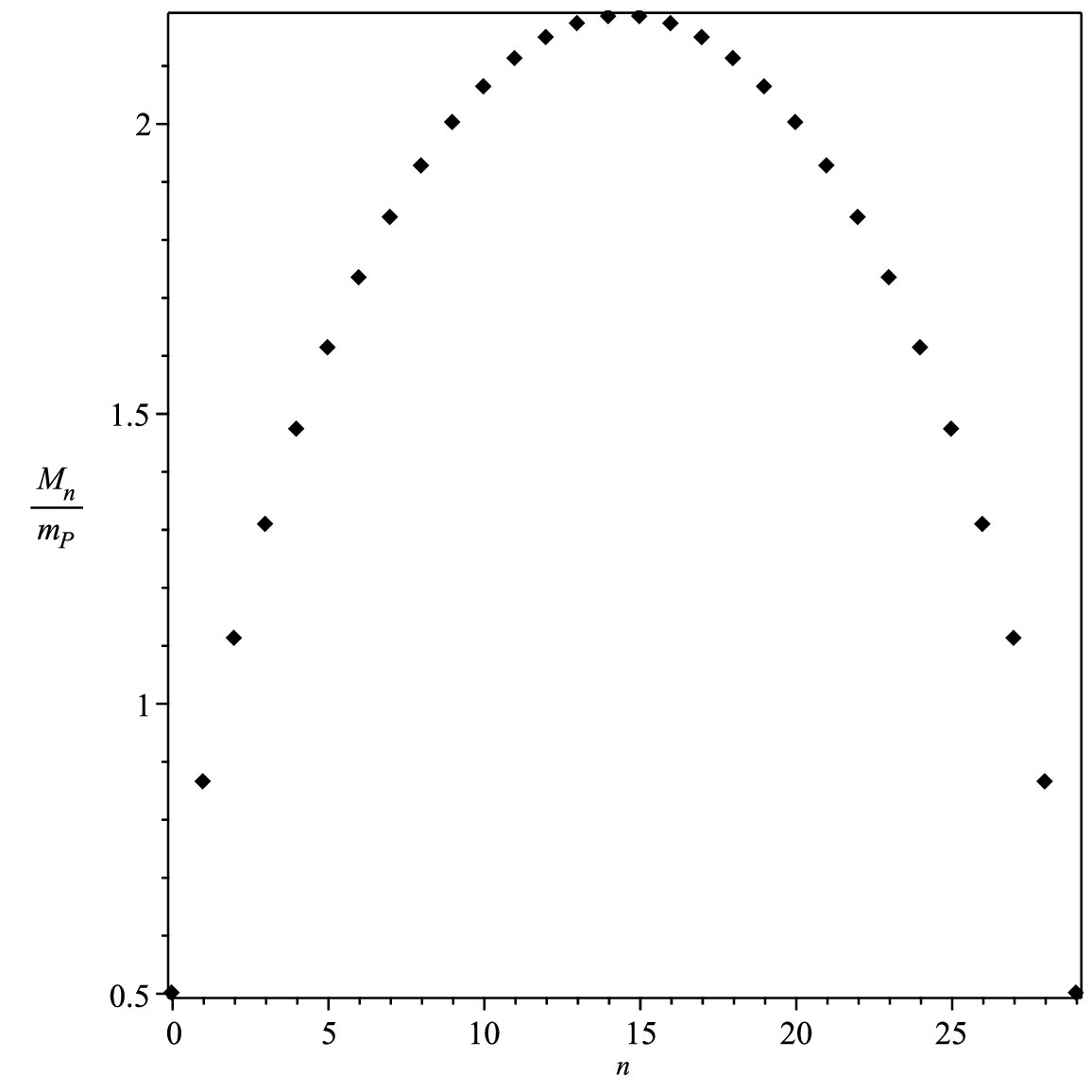}
       \caption{The mass spectrum of a $q$-deformed BH-WH pair.}
         \label{Fig1}
     \end{subfigure}
     \hfill
     \begin{subfigure}[b]{0.4\textwidth}
         \centering
         \includegraphics[width=\textwidth]{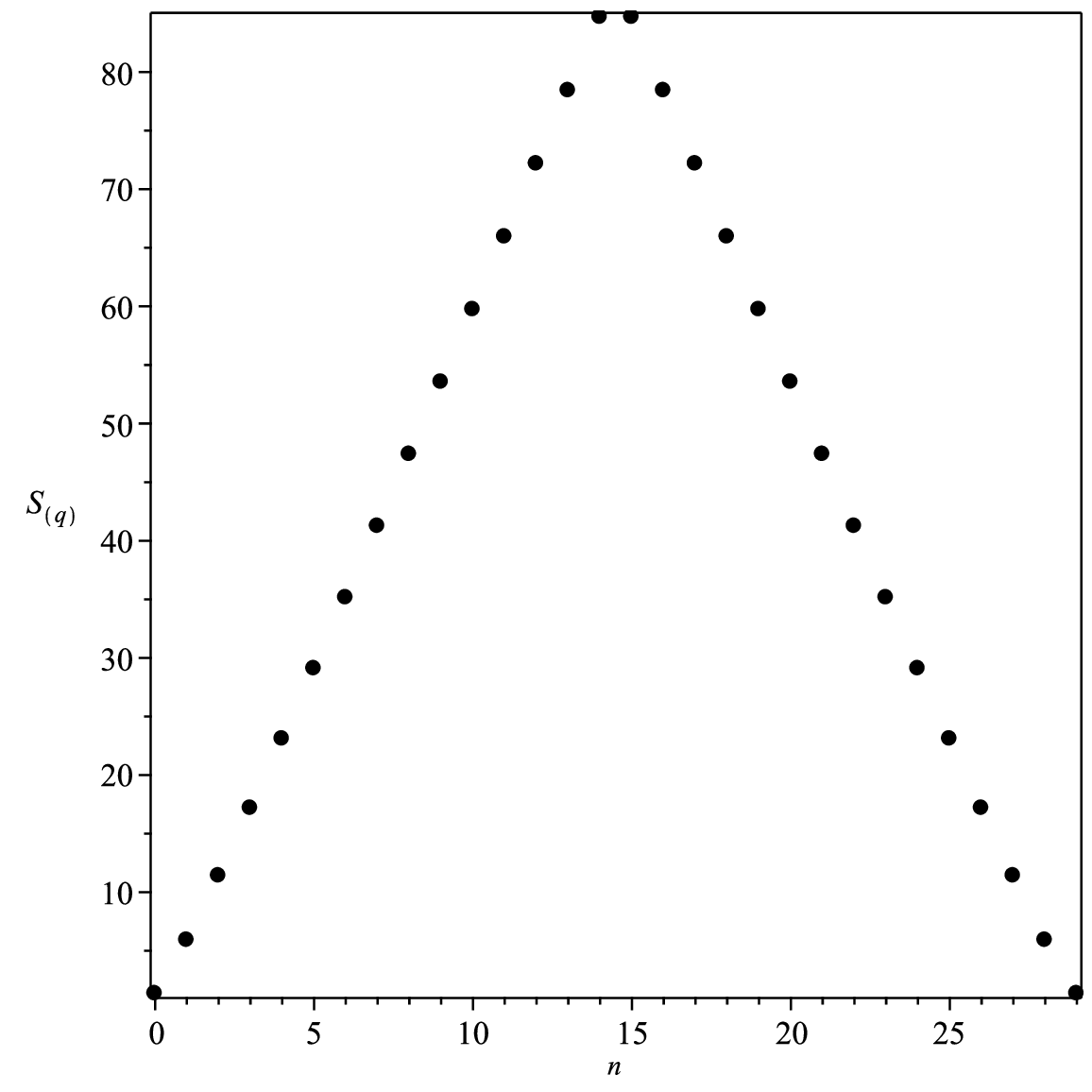}
         \caption{The entropy spectrum of a $q$-deformed BH-WH pair.}
         \label{Fig2}
     \end{subfigure}
            \caption{Fig. (a) shows the mass spectrum of a $q$-deformed BH-WH with $\mathfrak N=30$. Fig. (b) illustrates the entropy of the same BH-WH pair.}
        \label{Fig}
\end{figure}

Let us assume $\mathfrak N$ is an even number. {In the finite spectrum, the sector \(0 \leq n \leq \frac{N}{2}-1\) is identified as the BH branch because the mass increases with \(n\), so downward transitions \(n \to n-1\) correspond to mass loss, as expected for Hawking emission. The sector \(\frac{N}{2} \leq n \leq N-1\) is correspondingly interpreted as a WH branch, since the same bounded spectrum is traversed beyond the maximal-mass states and \(\frac{dM_n}{dn}<0\). We stress that `BH' and `WH' here refer to the two monotonic branches of a single minisuperspace spectrum, not to a literal two-body astrophysical system \cite{Jalalzadeh:2022rxx}.}  States with maximum mass eigenvalue $M_\text{max}=\frac{m_\text{P}}{2}\sqrt{\cot(\frac{\pi}{2\mathfrak N})}$ are $n=\mathfrak N/2-1$ and $n=\mathfrak N/2$. To illustrate this point, Fig. \ref{Fig1} shows the mass spectrum of a $q$-deformed BH with $\mathfrak N=30$. 
Using (\ref{3-19non1}) one checks
$\frac{\partial M_n}{\partial n}\propto \cos\!\big(\tfrac{\pi}{\mathfrak N}(n+\tfrac12)\big)$,
hence $M_n$ increases for $0\le n<\mathfrak N/2$ (BH branch) and decreases for $n>\mathfrak N/2$ (WH branch).
The stationary points at $n=\mathfrak N/2-1,\ \mathfrak N/2$ are maxima and yield
$M_{\max}=\frac{m_\text{P}}{2}\sqrt{\cot(\frac{\pi}{2\mathfrak N})}$, confirming the twofold degeneracy.
Note that $dM/dn>0$ for $n<\mathfrak N/2$ and $<0$ for $n>\mathfrak N/2$.
In addition, the wavefunctions are q-Hermite polynomials whose properties can be found in Refs. \cite{Jalalzadeh:2017jdo, 1994PhLB150B}.

\section{Thermodynamics of $q$-deformed black and white hole pair}
We assume Hawking radiation from a massive BH (WH) emitted during a transition from state \( n \) to state \( n-1 \) (from state \( n-1 \) to state \( n \)). Also, we consider that \( M \gg m_\text{P} \), \( n \gg 1 \), and $\mathfrak N\gg2$, which implies that $\sin(\frac{\pi}{2\mathfrak N})=\tan(\frac{\pi}{2\mathfrak N})\simeq\frac{\pi}{2\mathfrak N}$. It has been established \cite{Area1a, Xiang} that the dynamics of the BH occur between discrete mass eigenstates given by Eq. (\ref{3-19non1}). Then, using the $q$-deformed mass spectrum (\ref{3-19non1}) gives us the frequency of emitted (or absorbed) radiation \cite{Jalalzadeh:2025uuv}.

For adjacent levels, $|\Delta M|=|M_n-M_{n-1}|\simeq \big|\frac{\partial M}{\partial n}\big|$ (with $\Delta n=1$).
Differentiating (\ref{3-19non1}) w.r.t.\ $n$ and expanding for large $\mathfrak N$ yields
\(
\frac{\partial M}{\partial n}\simeq \frac{m_\text{P}}{8 M_n}\Big[1+\frac{m_\text{P}^2}{8M_n^2}-\frac{1}{2}\big(\frac{2\pi M_n^2}{\mathfrak N m_\text{P}^2}\big)^2\Big].
\)
Using $\omega=\frac{|\Delta M|}{2\pi}$ and $t_\text{P}=1/m_\text{P}$ ensures dimensional consistency in natural units,
\begin{equation}\label{4-1} \omega=\frac{|\Delta M|}{2\pi}\simeq \frac{m_\text{P}\sqrt{1-\left(\frac{2\pi M_n^2}{\mathfrak Nm_\text{P}^2} \right)^2} }{8\pi M_nt_\text{P}}\left(1+\frac{m_\text{P}^2}{8M_n^2}\sqrt{1-\left(\frac{2\pi M_n^2}{\mathfrak Nm_\text{P}^2} \right)^2}  \right),
\end{equation}
  where $|\Delta M|=|M_{n}-M_{n-1}|$, and $t_\text{P}$ is the Planck time. Note that at the limit $\mathfrak N\rightarrow\infty$, the above  relation reduces to
   \begin{equation}\label{3-1}
   \omega\simeq\frac{ m_\text{P}}{8\pi M_nt_\text{P}}\left(1+\frac{1}{8}\left(\frac{m_\text{P}}{M_n}\right)^2\right),~~~~n\gg1, \end{equation}
  which is the frequency of the radiation obtained in ordinary BHs \cite{Jalalzadeh:2025uuv}.

The following adiabatic invariant of the model, as demonstrated in Refs. \cite{Jalalzadeh:2025uuv, Kunstatter:2002pj, Shahjalal:2019ypz} for ordinary BHs, gives the entropy of a $q$-deformed BH (WH):
\begin{equation}
    \label{BHI1}
S_{(q)}=\int_{m_\text{P}/2}^M\frac{\mathrm dM}{\omega}\simeq2\mathfrak N\arcsin\left( \frac{S_\text{BH}}{2\mathfrak N}\right)-\frac{\pi}{2}\ln\left({S_\text{BH}}\right)+\text{const.},
\end{equation}
where $S_\text{BH}=4\pi M^2/m_\text{P}^2$ is the Bekenstein--Hawking entropy of an ordinary Schwarzschild BH. {A brief derivation of Eq.~(\ref{BHI1}), showing explicitly how the adiabatic-invariant integral yields both the arcsine contribution and the logarithmic correction, is presented in \ref{app:entropy}}.  Also, by assuming $\frac{S_\text{BH}}{2\mathfrak N}\ll1$, and using the Taylor series expansion of $\arcsin(x)$, the entropy (\ref{BHI1}) gives us the cubic correction to the hole entropy:
 \begin{equation}
     S_{(q)}\simeq S_\text{BH}+\frac{1}{24\mathfrak N^2}\left(S_\text{BH}\right)^3-\frac{\pi}{2}\ln\left(S_\text{BH}\right).
 \end{equation}
Note that for \(\mathfrak{N} \rightarrow \infty\), the aforementioned $q$-deformed entropy converges to the Hawking--Bekenstein entropy with a logarithmic correction, where the coefficient $-\pi/2$ arises from the modified state density in the $q$-deformed measure, distinguishing it from the LQG value $-3/2$. term:
\begin{equation}
 \label{BHI}
S=S_\text{BH}-\frac{\pi}{2}\ln\left(S_\text{BH}\right)+\text{const.}
 \end{equation}
 Therefore, the \(q\)-deformation generates a negative logarithmic correction to the entropy, in addition to the bounded arcsine contribution associated with the finite-dimensional spectrum.
 It is worth focusing on the logarithmic term obtained here
($\ln(S_{BH})$). Previous studies insist on the universality of
this term
\cite{Rovelli:1996dv,Ashtekar:1997yu,Solodukhin:1997yy,Mann:1997hm,
Kaul:2000kf,Ghosh:2004rq,Domagala:2004jt,Meissner:2004ju,Hod:2004di,
Medved:2004eh,Kastrup:1997iu,Das:2001ic,Cai:2008ys}.
Therefore, the result indicates that just like the frameworks of
quantum geometry of the horizon
\cite{Rovelli:1996dv,Ashtekar:1997yu,Solodukhin:1997yy,Mann:1997hm,
Kaul:2000kf,Ghosh:2004rq,Domagala:2004jt,Meissner:2004ju} and
thermodynamic fluctuations around the equilibrium
\cite{Hod:2004di,
Medved:2004eh,Kastrup:1997iu,Das:2001ic,Mukherji:2002de,
Gour:2003jj, Chatterjee:2003uv, Cai:2008ys}, the $q$-deformed algebra also requires this term.
 
Furthermore, comparing the differential of (\ref{BHI1}) with the first law of thermodynamics for BHs, \(\mathrm dM=T_\text{H}\mathrm dS\), provides the temperature of the $q$-deformed BH-WH pair:
\begin{equation}
    \label{Temp}
        T_{(q)}=\frac{m_\text{P}}{8\pi M}\sqrt{1-\left(\frac{2\pi M^2}{\mathfrak Nm_\text{P}^2}\right)^2}\left\{1+\frac{m_\text{P}^2}{8M^2}\sqrt{1-\left(\frac{2\pi M^2}{\mathfrak Nm_\text{P}^2}\right)^2} \right\} T_\text{P},
\end{equation} 
where $T_\text{P}$ is the Planck temperature. It is easy to verify that the combination of Eqs. (\ref{Temp}) and (\ref{BHI1}), and ignoring the logarithmic term, gives us
\begin{equation}
    \label{Smarr}
 M=4\mathfrak N T_\text{(q)}\tan\left(\frac{S_\text{{(q)}}}{2\mathfrak N}\right).
\end{equation}
{This equation is the $q$-deformed Smarr’s formula for the mass of a BH. In the limit $\mathfrak N\to\infty$,
$\tan(S_{(q)}/(2\mathfrak N))\to S_\text{BH}/(2\mathfrak N)$ and the standard Smarr relation $M=2T_\text{H}S_\text{BH}$ \cite{Smarr:1972kt} is recovered. It is clear that for \(\mathfrak{N} \rightarrow \infty\), this relation reduces to Smarr’s formula $M=2T_\text{H}S_\text{BH}$.}

 Also, the heat capacity is given by
 \begin{equation}\label{Heat}
C_{(q)}=\frac{dM}{dT_{(q)}}=-\frac{8\pi M^2\sqrt{1-\left(\frac{2\pi M^2}{\mathfrak Nm_\text{P}^2}\right)^2} }{m_\text{P}^2\left({1+\left(\frac{2\pi M^2}{\mathfrak Nm_\text{P}^2}\right)^2} \right)}.
 \end{equation}
One can easily find out that the above temperature and heat capacity reduce to the Hawking temperature, $T_H=m_\text{P}/(8\pi M)T_\text{P}$, and the heat capacity, $C=-8\pi M^2/m_\text{P}^2$, of the Schwarzschild hole at $\mathfrak N\rightarrow\infty$ limit.

{As in the Schwarzschild case, Hawking emission drives the BH branch toward lower mass. The negativity of the heat capacity in Eq. (\ref{Heat}) implies that this mass loss is accompanied by an increase in temperature along the branch.} On the other hand, the WH radiates to the insides of its horizon, and it gains more mass until its mass reaches the maximum mass $M_\text{max}=\frac{m_\text{P}}{2}\sqrt{\cot(\frac{\pi}{2\mathfrak N})}\simeq m_\text{P}\sqrt{\mathfrak N/(2\pi)}$. To clarify these principles, we shall examine multiple virtual particle pairs that arise in the vicinity of the outer region of the horizon. The intense gravitational field applies a stronger attractive force on the particle that is closer in proximity compared to the one that is farther away; therefore, the gravitational pull of the BH generates a tidal force that acts to dissociate the virtual pair. The substantial force present near the event horizon culminates in the irreversible separation of particles, resulting in their conversion into real particles, as they are no longer able to reunite for mutual annihilation. The particle that is closest to the horizon is engulfed, while the one that is more distant is allowed to escape to infinity due to the energy transferred by the tidal force. Consequently, there is a decrease in the mass of the BH. In contrast, the conditions surrounding a WH are opposed to the previously discussed situation. If we consider the creation of virtual particles within the horizon of a WH, the particle crosses the horizon and is unable to re-enter the interior of the horizon. In the case of a WH, it gains mass through the process of Hawking radiation.

 Note that inserting the mass spectrum (\ref{3-19non1}) into (\ref{BHI1}), and ignoring the logarithmic term, gives us the discrete $q$-deformed entropy of a massive BH-WH pair as
\begin{equation}
    \label{spec1}
    S_{(q)}=\begin{cases}2\pi\left(n+\frac{1}{2}\right),\hspace{2.3cm}0\leq n\leq\frac{\mathfrak N}{2}-1,~~~\text{(BH)},\\
        2\pi\left(\mathfrak N-n-\frac{1}{2}\right),\hspace{1.33cm}\frac{\mathfrak N}{2}\leq n\leq\mathfrak N-1, ~~~\text{(WH)}.
    \end{cases}
\end{equation}
This entropy relation is consistent with the Bekenstein proposal \cite{1974NCimL}, given by Eq. (\ref{Ent}). However, we should note that the $q$-deformed entropy is bounded from above. To illustrate this point,
Fig. \ref{Fig2} shows the entropy spectrum of a $q$-deformed BH-WH pair for $\mathfrak N=30$. The maximum possible value for entropy (for $n=\mathfrak N/2-1$ or $n=\mathfrak N/2$) is
\begin{equation}
    \label{Maxentropy}
    S_{(q)}^\text{(max)}=\pi(\mathfrak N-1)\simeq\pi\left(\frac{L_q}{l_\text{P}} \right)^2.
\end{equation}
Therefore, the WH gains more mass via Hawking radiation and achieves its maximum entropy given by the above relation. Note that the existence of maximum entropy and also maximum mass, given by Eq. (\ref{3-19non1}) are direct results of $q$-deformation. In addition, the temperature and the heat capacity of this state are given by
\begin{equation}
    \label{Maxtem}
    T_{(q)}^\text{(min)}\simeq\sqrt{\frac{\pi}{128}}\left(\frac{l_\text{P}}{L_q}\right)^3,~~~~C_\text{(q)}^\text{(max)}\simeq-\pi.
\end{equation}
Eqs. (\ref{Maxentropy}) and (\ref{Maxtem}) demonstrate that the ultimate state of WHs is universally applicable, irrespective of the initial mass of the WH, and is wholly a manifestation of quantum gravitational phenomena.

{We emphasize that the analysis of Sections~2--3 is performed for the \(q\)-deformed minisuperspace quantization of the Schwarzschild geometry, not for a Schwarzschild--de Sitter metric. The quantity \(\Lambda_q = 3/L_q^2\) enters only as an effective large-scale parameter inferred from the maximal entropy \(S_q^{\max}\), which coincides formally with the de Sitter entropy. Our analysis does not quantize a Schwarzschild--de Sitter geometry. Thus, the de Sitter relation $S_{(q)}^\text{(max)}=3\pi/(G\Lambda_q)$ should be viewed as an effective holographic/cosmological correspondence implied by the finite-dimensional $q$-deformed Hilbert space, rather than as an assumption about the asymptotics of the black-hole spacetime studied in Sections 2 and 3.} 
{Also, at the algebraic level, the key input is simply that $q$ is a root of unity, which makes the representation finite dimensional with dimension $\mathfrak N$. All spectral boundedness results in Eqs. (\ref{3-19non1}), (\ref{Maxentropy}), and (\ref{Maxtem}) follow from finite $\mathfrak N$ alone. The parametrization $\mathfrak N=L_q^2/l_\text{P}^2$ is an additional physical identification that introduces an infrared length scale $L_q$; only after this identification does one obtain the effective cosmological quantity $\Lambda_q=3/L_q^2$.}


\section{Conclusions}

 Quantum groups provide more intricate symmetries than conventional Lie algebras, of which they are a special subset.  This suggests that quantum groups may be suitable for characterizing the symmetries of physical systems that exceed the limitations of Lie algebras.  Furthermore, $q$-deformed models provide a notable benefit due to their associated Hilbert space being finite-dimensional when $q$, the deformation parameter, is a root of unity \cite{Chaichian2}.  This indicates that using quantum groups with a deformation parameter of the root of unity is advantageous for developing models with a limited number of states.  These models can be used to investigate applications in quantum gravity and quantum cosmology that conform to the holographic principle and UV/IR mixing, addressing the cosmological constant problem \cite{Jalalzadeh:2017jdo}. In a more explicit manner, within the framework of cosmology, the quantum deformation parameter $L_q$, which is articulated subsequent to the commutation relations (\ref{new4}), results in the emergence of a cosmological constant represented as $\Lambda_q=3/L_q^2$.
 Identifying $\mathfrak N=L_q^2/l_\text{P}^2$ implies $S_{(q)}^{\max}=\pi(L_q/l_\text{P})^2=\frac{3\pi}{G\Lambda_q}$, equal to the de Sitter entropy for $\Lambda_q$, thus linking the IR deformation scale to cosmology.

{In this work, we studied the thermodynamic implications of the \(q\)-deformed minisuperspace quantization of the Schwarzschild black hole. The root-of-unity deformation leads to a finite-dimensional Hilbert space and, consequently, to a bounded mass spectrum with two monotonic branches, which we interpret as black-hole and white-hole sectors of a single quantum spectrum. Using the adiabatic-invariant method, we obtained a modified entropy formula containing both a bounded arcsine term and a logarithmic correction. The corresponding temperature and heat capacity exhibit a minimum temperature and a maximal entropy, suggesting an effective infrared cutoff scale. The relation of this cutoff to an effective cosmological constant was discussed at the thermodynamic level. Further investigation is needed to determine whether these features persist beyond the present minisuperspace framework and whether they admit observable phenomenological consequences.}

Potential, albeit challenging, signatures include faint gravitational-wave echoes from transitions between adjacent $q$-levels. Extending the analysis to Reissner--Nordström and Kerr--Newman geometries is a priority for astrophysical relevance.

\appendix
\section{Canonical reduction of the Schwarzschild black hole and the oscillator form of the WDW equation}\label{canon}

In this appendix, we briefly review the canonical reduction of the Schwarzschild BH that underlies the WDW equation used in the main text. Since Eq.~(\ref{WDW1}) plays a central role in our analysis, it is useful to summarize how it arises from the reduced phase-space quantization of spherically symmetric gravity.

We begin with the spherically symmetric ADM line element
\begin{equation}\label{1-1}
ds^2=-N(r,t)^2dt^2+\Lambda(r,t)^2\Big(dr+N^r(r,t)dt\Big)^2+R(r,t)^2d\Omega^2,
\end{equation}
where \(d\Omega^2\) is the line element on the unit two-sphere \(S^2\). We adopt Kucha\v r's fall-off conditions \cite{Kuchar1}, which ensure that the coordinates \(r\) and \(t\) extend over the Kruskal manifold, \(-\infty<r,t<\infty\), while the geometry remains asymptotically flat at both spatial infinities. These conditions also imply that the ADM 4-momentum has no spatial component at \(r\rightarrow\pm\infty\), so that the BH is at rest with respect to the left and right asymptotic Minkowski regions.

Fixing the asymptotic values of the lapse function at the two infinities to be time-dependent functions \(N_\pm(t)\), the Einstein--Hilbert action, supplemented by the appropriate boundary terms, takes the Hamiltonian form
\begin{equation}\label{1-2}
\mathcal S=\int dt\int_{-\infty}^{\infty}\Big\{ \Pi_\Lambda \dot\Lambda +\Pi_R\dot R-NH-N^rH_r\Big\}dr -\int\Big\{N_+M_++N_-M_-\Big\}dt,
\end{equation}
where the canonical momenta conjugate to \(\Lambda\) and \(R\) are
\begin{equation}\label{1-3}
\begin{split}
\Pi_\Lambda&=-\frac{m_\text{P}^2}{N}R\Big(\dot R-R'N^r\Big),\\
\Pi_R&=-\frac{m_\text{P}^2}{N}\Big[\Lambda\left(\dot R-R'N^r\right)+R\left(\dot\Lambda-(\Lambda N^r)'\right)\Big].
\end{split}
\end{equation}
Here \(m_\text{P}=1/\sqrt{G}\) is the Planck mass, \(\dot f=\partial_t f\), and \(f'=\partial_r f\). The quantities \(M_\pm(t)\) are determined by the asymptotic fall-off of the canonical variables and, on classical solutions, coincide with the Schwarzschild mass.

The super-Hamiltonian and radial super-momentum constraints are given by
\begin{equation}\label{new1}
\begin{split}
H&=-\frac{1}{Rm_\text{P}^2}\Pi_R\Pi_\Lambda+\frac{1}{2R^2m_\text{P}^2}\Pi_\Lambda^2+\frac{RR''}{\Lambda}
-\frac{RR'\Lambda'}{\Lambda^2}+\frac{R'^2}{2\Lambda}-\frac{\Lambda}{2},\\
H_r&=\frac{1}{m_\text{P}^2}\left(\Pi_RR'-\Lambda\Pi'_\Lambda\right).
\end{split}
\end{equation}

Following Kucha\v r \cite{Kuchar1}, one introduces the canonical transformation from the original variables \((\Lambda,\Pi_\Lambda;R,\Pi_R)\) to the new pairs \((M,\Pi_M)\) and \((\mathcal R,\Pi_{\mathcal R})\),
\begin{equation}\label{new2}
\begin{split}
M&=\frac{\Pi_\Lambda^2}{2m_\text{P}^4R}-\frac{RR'^2}{2\Lambda^2}+\frac{R}{2},\\
\Pi_M&=\frac{\Lambda\Pi_\Lambda}{m_\text{P}^2}\left[\left(\frac{R'}{\Lambda}\right)^2-\frac{1}{m_\text{P}^4}\left(\frac{\Pi_\Lambda}{R}\right)^2 \right]^{-1},\\
\mathcal R&=R,\\
\Pi_{\mathcal R}&=\left(\frac{\Pi_\Lambda H}{m_\text{P}^2R}+\frac{R'H_r}{\Lambda^2} \right)\left[\left(\frac{R'}{\Lambda}\right)^2-\frac{1}{m_\text{P}^4}\left(\frac{\Pi_\Lambda}{R}\right)^2 \right]^{-1}.
\end{split}
\end{equation}
In these variables, the action becomes
\begin{equation}\label{new3}
\mathcal S=\int dt\int_{-\infty}^\infty\left\{ \dot M\Pi_M+\dot{\mathcal R}\Pi_{\mathcal R}-N^rH_r-NH\right\}dr
-\int \left\{M_+N_+-M_-N_-\right\}dt,
\end{equation}
where the transformed constraints take the form
\begin{equation}\label{new4d}
\begin{split}
H&=-\frac{\left(1-\frac{2M}{m_\text{P}^2R}\right)^{-1}M'{\mathcal R}'+m_\text{P}^{-4}\left(1-\frac{2M}{m_\text{P}^2R}\right)\Pi_M\Pi_{\mathcal R}}{\left[\left(1-\frac{2M}{m_\text{P}^2R}\right)^{-1}{\mathcal R}'^2-m_\text{P}^{-4}\left(1-\frac{2M}{m_\text{P}^2R}\right)\Pi_M^2 \right]^{1/2}},\\
H_r&=\frac{1}{m_\text{P}^2}\left(\Pi_MM'+\Pi_{\mathcal R}{\mathcal R}'\right).
\end{split}
\end{equation}

Variation of the action with respect to \(N\) and \(N^r\) imposes the constraints
\begin{equation}\label{new5}
H\approx0,\qquad H_r\approx0,
\end{equation}
which are equivalent to
\begin{equation}\label{new6}
M'\approx0,\qquad \Pi_{\mathcal R}\approx0.
\end{equation}
The first relation implies that \(M\) is independent of the radial coordinate, namely \(M=M(t)\). After imposing \(\Pi_{\mathcal R}\approx0\) and \(M=M(t)\), one obtains a reduced action in terms of the single canonical pair \((M,P)\), where
\begin{equation}\label{1-4}
P=\int_{-\infty}^\infty \Pi_M\,\mathrm dr=
-\int_{-\infty}^\infty\frac{\sqrt{\left(\frac{dR}{\mathrm dr}\right)^2-\Lambda\left(1-\frac{2M}{m_\text{P}^2R}\right)}}{1-\frac{2M}{m_\text{P}^2R}}\,\mathrm dr,
\qquad -\infty<P<\infty.
\end{equation}
The reduced action then takes the form
\begin{equation}\label{1-5}
\mathcal S=\int \Big\{P\dot M-(N_++N_-)M\Big\}\mathrm dt.
\end{equation}

The variables \((M,P)\) satisfy the canonical Poisson bracket \(\{M,P\}=1\). Choosing, as in Ref.~\cite{Louko}, the asymptotic Minkowski time on the right-hand side as the evolution parameter amounts to fixing
\begin{equation}
N_+=1,\qquad N_-=0,
\end{equation}
so that the reduced action becomes
\begin{equation}\label{1-6}
\mathcal S=\int \left\{P\dot M-H(M)\right\}\mathrm  dt,
\end{equation}
with reduced Hamiltonian
\begin{equation}
H(M)=M.
\end{equation}
The corresponding equations of motion yield \(M=\text{const.}\) and \(P=-t\), in agreement with Birkhoff's theorem: the Schwarzschild mass is the only gauge-invariant constant of motion of the vacuum solution. The conjugate momentum \(P\) may therefore be interpreted as the asymptotic time separation associated with the spatial hypersurface.

To proceed toward quantization, one follows the Euclidean argument of Ref.~\cite{Louko}. Since the Euclidean Schwarzschild geometry is regular only if the Euclidean time is periodic with period \(T_H^{-1}\), where
\begin{equation}
T_H=\frac{1}{8\pi GM},
\end{equation}
one imposes the identification
\begin{equation}\label{Time}
P\sim P+\frac{1}{T_H}.
\end{equation}
This periodicity removes the conical singularity at the horizon in the Euclidean section. At the same time, it shows that the physical phase space is not the full \((M,P)\)-plane, but rather a wedge bounded by the \(M\)-axis and the line \(P=T_H^{-1}\).

It is therefore convenient to introduce a new canonical pair \((x,p)\) that unwraps this wedge-shaped phase space and incorporates the periodic identification naturally:
\begin{equation}\label{Trans}
x=2\sqrt{G}\,M\cos(2\pi T_HP),\qquad
p=2\sqrt{G}\,M\sin(2\pi T_HP).
\end{equation}
Using \(\{M,P\}=1\), one readily verifies that this transformation is canonical, \(\{x,p\}=1\). Moreover, squaring and adding the two relations in Eq.~(\ref{Trans}) gives
\begin{equation}
M^2=\frac{m_P^2}{4}\left(x^2+p^2\right).
\end{equation}
Thus, in the reduced phase space, the Schwarzschild mass is mapped to the Hamiltonian of a one-dimensional harmonic oscillator. Upon canonical quantization in the \(x\)-representation,
\begin{equation}
p\rightarrow -i\frac{\mathrm d}{\mathrm dx},
\end{equation}
one obtains
\begin{equation}
\left[-\frac{1}{2}\frac{\mathrm d^2}{\mathrm dx^2}+\frac{1}{2}x^2\right]\psi(x)
=2\frac{M^2}{m_P^2}\psi(x),
\end{equation}
which is precisely Eq.~(\ref{WDW1}) in the main text.

We therefore interpret \(x\) as the dimensionless canonical minisuperspace variable of the reduced Schwarzschild BH, in terms of which the WDW equation takes the form of a harmonic-oscillator eigenvalue problem.

\section{Derivation of the entropy formula}
\label{app:entropy}

In this appendix, we provide an intermediate derivation of Eq.~(\ref{BHI1}), showing explicitly how the adiabatic-invariant integral yields both the arcsine term and the logarithmic correction.

Starting from Eq.~(\ref{4-1}), we treat the mass spectrum semiclassically and regard \(M_n\) as a continuous variable \(M\). The inverse of the transition frequency can then be written as
\begin{equation}
\label{B1}
\begin{split}
\frac{1}{\omega(M)}&=\frac{{8\pi M}}{m_\text{P}^2\sqrt{1-\left(\frac{2\pi M^2}{\mathfrak N m_\text{P}^2}\right)^2}}
\left[
1+\frac{m_\text{P}^2}{8M^2\sqrt{1-\left(\frac{2\pi M^2}{\mathfrak N m_P^2}\right)^2}}
\right]^{-1}\\
&\simeq-\frac{\pi}{M}+\frac{8\pi M}{m_\text{P}^2\sqrt{1-\left(\frac{2\pi M^2}{\mathfrak N m_\text{P}^2}\right)^2}},
\end{split}
\end{equation}
where, in the second equality, we did expand \(1/\omega(M)\) up to the first subleading order.
 The corresponding adiabatic invariant is
\begin{equation}
\label{B2}
I=\int \frac{\mathrm dM}{\omega(M)}\simeq
\int
\frac{8\pi M\,\mathrm dM}{m_\text{P}^2\sqrt{1-\left(\frac{2\pi M^2}{\mathfrak N m_\text{P}^2}\right)^2}}
-\pi\int \frac{\mathrm dM}{M}.
\end{equation}

The first integral gives 
\begin{equation}
\label{B6}
\int
\frac{8\pi M\,\mathrm dM}{m_\text{P}^2\sqrt{1-\left(\frac{2\pi M^2}{\mathfrak N m_\text{P}^2}\right)^2}}
=
2\mathfrak N  \arcsin\!\left(\frac{2\pi M^2}{\mathfrak N m_\text{P}^2}\right).
\end{equation}
The second integral is elementary:
\begin{equation}
\label{B7}
-\pi\int \frac{\mathrm dM}{M}
=
-\pi\ln M+\text{const.}
\end{equation}
Using the Bekenstein--Hawking entropy
\begin{equation}
\label{B8}
S_\text{BH}=\frac{4\pi M^2}{m_\text{P}^2},
\end{equation}
one has
\begin{equation}
\label{B10}
-\pi\ln M
=
-\frac{\pi}{2}\ln S_{BH}+\text{const.}
\end{equation}

Combining the two contributions, we find the result,
\begin{equation}
\label{B11}
S_q\simeq
2\mathfrak N\arcsin\!\left(\frac{S_\text{BH}}{2\mathfrak N}\right)
-\frac{\pi}{2}\ln S_\text{BH}
+S_0,
\end{equation}
which is Eq.~(\ref{BHI1}) of the main text.

Thus, the arcsine term originates from the integral of the deformed square-root factor in the transition frequency, whereas the logarithmic correction arises from the subleading term in the expansion of \(1/\omega(M)\).

\vspace{.3cm}
\section*{References}
\bibliographystyle{iopart-num}       
\bibliography{Qdeformed}   


\end{document}